\documentclass[12pt]{article}
\usepackage{cite}
\usepackage{preprint2}

\preprint{DESY--01--013\\
          physics/0102024
}

\title{On the Compatibility Between Physics and Intelligent Organisms}

\author{%
   John C. Collins,\footnote{E-mail: collins@phys.psu.edu}
   \footnote{On leave from:
        Physics Department,
        Penn State University, 
        104 Davey Laboratory,
        University Park PA 16802,
        U.S.A.
   }
   \\
   DESY, Notkestra{\ss}e 85, D-22603 Hamburg, Germany, \\
   {\em and}\\
   II Institut f{\"u}r Theoretische Physik, Universit{\"a}t Hamburg, \\{}
   Luruper Chaussee 149, D-22761 Hamburg, Germany
}

\date{9 February 2001}

\begin{document}
\maketitle

\begin{abstract}
It has been commonly argued, on the basis of G{\"o}del's theorem and
related mathematical results, that true artificial intelligence cannot
exist.  Penrose has further deduced from the existence of human
intelligence that fundamental changes in physical theories are needed.
I provide an elementary demonstration that these deductions are
mistaken.
\end{abstract}


Is real artificial intelligence possible? Are present-day theories of
physics sufficient for a reductionist explanation of consciousness? Among
the long history of discussions of these questions
\cite{Lucas,Penrose,BBS,Psyche}, the eloquent writings of Penrose
\cite{Penrose} stand out for strong mathematical arguments that give
negative answers to both questions.

For a physicist, Penrose's result is quite striking.  He claims that
understanding the human brain entails big changes in current microscopic
theories of physics (e.g., quantum mechanics in general and quantum
gravity in particular). This is contrary to our normal scientific
experience. The enormous progress in both elementary particle physics and
in molecular biology during the last 30 years has had no direct mutual
influence on the two fields, except on our students' career choices. Now
we have an eminent theoretical physicist telling elementary particle
physicists to look to neurons rather than multi-billion-dollar
accelerators for progress in physics. Should we believe him?

In this paper, I provide an elementary argument that this chain of
reasoning fails.

\section{Penrose's argument}

Penrose observes that current microscopic theories of physics are
computational, and that they appear to underlie all chemical and
biological phenomena. It follows that it is possible to simulate all
properties of biological organisms by a computer program. Of course, the
computer program will be impractically big to implement, but this fact
does not affect Penrose's mathematics. Among the biological organisms are
mathematicians, so the computer program provides artificial mathematicians
that are completely equivalent to human mathematicians.  This runs afoul
of Turing's halting theorem, which, taken at face value, implies that
artificial mathematicians are always less powerful than human
mathematicians.

{}From this contradiction, Penrose deduces that better theories of physics
are needed and that the new theories must be non-computational, unlike
current theories, such as the ``Standard Model'', which are all particular
quantum mechanical theories.  On the way he also demolishes all hope for
true artificial intelligence.

Of course, this argument attracted much comment \cite{BBS,Psyche}. The
critics observe that real computer programs appear to have a much richer
range of behavior than the kind of computation used in Turing's theorem.
This theorem applies to strictly non-interactive computer programs
(technically known as Turing machines), whereas real intelligent entities
are obviously much more like interactive computer programs.  But Penrose
always appears to have a comeback. For example, if an intelligent computer
needs to be trained by an environment, then he tells us to simulate the
environment by a computer, just as one might hook up an aircraft control
computer to a flight simulator instead of a real crashable jumbo jet.

Thus Penrose counters the criticism by observing that from an interactive
program one can construct a non-interactive program, i.e., one that does
not depend on repeated interaction with other beings or with an
environment.  I will show that this construction fails.  The construction
of a non-interactive program satisfying a particular precise specification
of the kind needed in the Turing theorem inevitably loses access to the
full powers of a putative intelligent interactive program.

\section{Turing's halting theorem}

The technical results use the concept of a ``Turing machine''. Now a Turing
machine is simply any device for performing a defined
computation. Turing's achievement was to characterize this mathematically,
i.e., to define in general what a computer program is. Hence, instead of
Turing machines, we can equally well discuss realistic computers and
actual programming languages.

The halting theorem --- see Penrose's excellent treatment \cite{Penrose}
--- concerns a subroutine $T_k(n)$ which takes one argument and which
obeys the following specification:
\begin{quote}
    $T_k(n)$ halts if and only if it has constructed a correct proof that
    the one-argument subroutine defined by $n$ does not halt when
    presented with data $n$.
\end{quote}
Here, the subscript in $T_k$
represents the source code for the subroutine. The argument $n$ is the
source code for another subroutine, and $T_k$ concerns itself with proving
properties of this second subroutine.

Turing's halting theorem is obtained when one sets $n=k$, i.e., when $T_k$
is asked to prove a theorem about itself. There is a contradiction unless
$T_k(k)$ does not halt; this result is exactly the halting theorem. From
the subroutine's specification, we see that the subroutine is unable to
prove this same theorem, the one stated in the previous sentence.

But humans can prove the theorem. From this follow the conclusions about
the impossibility of artificial intelligence, etc.

\section{Non-Turing computations }

What is the relation between a dry abstract theorem-proving subroutine and
a computer program simulating biological organisms? The behavior of the
organisms (even the mathematicians) is clearly a lot richer and more
varied than that of the theorem prover. Basically the answer is in the
common assertion that all computers and computer programs are examples of
Turing machines; that is, they can each be viewed as some subroutine
$T_k$. To obtain the Turing machine used in the halting theorem, one
simply has to ask the simulated mathematician to prove an appropriate
theorem.

However, the common assertion, of the equivalence between Turing machines
and computer programs, is not exactly correct. The idea of a Turing
machine is that it is given some definite input, it runs, and then it
returns the results of the computation. This is appropriate for the
calculation of a trigonometric function, for example. But a real computer
program may be interactive; it may repeatedly send output to the outside
world and receive input in response, word processors and aircraft control
programs being obvious examples. Such computer programs are not Turing
machines, strictly speaking. As Geroch and Hartle \cite{GH86} have
observed, a Turing machine is equivalent to a particular kind of computer
program, a program whose input is all performed in a single statement that
is executed once.

The legalistic distinction between Turing and non-Turing computations
matters critically for Penrose's results. Software that attempts to mimic
real intelligence must surely be in the more extended class of interactive
programs. Moreover, if one is to avoid programming all the details of its
behavior, the program must learn appropriate responses in interaction with
an environment. The prototypical case is unsupervised learning by an
artificial neural network, an idea with obvious and explicit inspiration
from biological systems.

To be able to using the halting theorem, one must demonstrate that, given
some software that genuinely reproduces human behavior, one can construct
from it a subroutine of the Turing type suitable for use in the theorem.

\section{Interactive programs}

Penrose \cite{Penrose} gives a number of examples, that appear to show
that it is easy to construct the requisite non-interactive subroutine
using the interactive program as a component.

However, there is a big problem in figuring out how to present the input
to the program, to tell it what theorem is to be proved. Now the program,
which we can call an artificial mathematician, is in the position of a
research scientist whose employer specifies a problem to be worked on. To
be effective, such a researcher must be able to question the employer's
orders at any point in the project. The researcher's questions will depend
on the details of the progress of the research. (``What you suggested
didn't quite work out. Did you intend me to look at the properties of XXYZ
rather than XYZ?'')  As every scientist knows, if the researcher does not
have the freedom to ask unanticipated questions, the whole research
program may fail to achieve its goals.

Therefore to construct the non-interactive program needed by Penrose one
must discover the questions the artificial mathematician will ask and
attach a device to present the answers in sequence.\footnote{
   In the case of a complete microscopic simulation of the real world,
   one must also figure out how to present mathematics research problems to
   the beings that are created by the simulation.  This is quite non-trivial
   given that the actual programming concerned itself exclusively with the
   interactions of quarks, gluons and electrons.  Nevertheless let us assume
   that this problem has been solved.
}
The combination of the original computer and the answering machine is the
entity to which Turing's halting theorem is to be applied.

How does one discover ahead of time the questions that will be asked?
(Remember that the program is sufficiently complex that one does not
design it by planning ahead of time the exact sequence of instructions to
be executed.) One obvious possibility is simply to run the program
interactively to discover the questions. Then one programs the answering
machine with the correct answers and reruns the program.

This is exactly what a software manufacturer might do to provide a
demonstration of a graphical design program. Both the graphical design
program and the answering machine are interactive programs; but the
combination receives no input from the outside world and is therefore an
actual Turing machine.

Here comes a difficulty that as far as I can see is unsolvable. The first
input to the program was a request to prove a particular theorem about a
particular computing system. This computing system happened to be the
program itself, {\em together with all its ancillary equipment}. When one
reruns the program after recording the answers to its questions, the
theorem under consideration has changed. The theorem is now about the
original computing system including the answering machine, and, most
importantly, the answers recorded on it.

The answers were recorded when the program was asked to prove a theorem
about the computing system with no answers on the answering machine. Why
should the questions remain the same when the theorem has changed? If they
don't, then the recorded answers can easily be wildly inappropriate.

Of course, the theorem has not changed very much. However, in a complex
computing system the output often depends sensitively on the details of
the input. Indeed, a system that is intended to be intelligent and
creative should show just such unpredictable behavior.

No matter how one goes about it, to discover the exact questions that the
artificial mathematician will ask requires us to know the answers. But one
doesn't know which answers are needed until one knows the questions. And
one must know the exact questions and answers, for otherwise one cannot
set up the subroutine used in Turing's halting theorem. The subroutine is
asked to prove a theorem $K$ about a certain subroutine. The proof of
Turing's halting theorem is inapplicable if even one bit of machine code
differs between the subroutine that attempts to prove the theorem $K$ and
the subroutine that is the subject of the theorem.

Once one realizes that the exact information on the questions and answers
cannot be found, the applicability of the halting theorem to a simulation
of biological organisms fails, and with it Penrose's chain of argument.

\section{Conclusion}

Intelligent software must behave much more like a human than the kinds of
software that are encompassed by the strict definition of a Turing
machine. Penrose's conclusion requires taking absolutely literally the
idea that every computation can be reduced to some Turing machine, so that
he can use Turing's halting theorem. The proof of the theorem requires
perfect equality between a certain subroutine that proves theorems and the
subroutine that is the subject of a theorem to be proved. But the
practicalities of converting intelligent software to the non-interactive
software used in the halting theorem preclude one from achieving this
exact equality.

We see here an example of a common phenomenon in science: any statement we
make about the real world is at least slightly inaccurate. When we employ
logical and mathematical reasoning to make predictions, the reasoning is
only applicable if it is robust against likely deviations between the
mathematics and the real world. This does not seem to be the case for
Penrose's reasoning.

\section*{Acknowledgments}
I would like to thank A. Ashtekar, S. Finn, J. Hartle, D.  Jacquette,
R. Penrose and L. Smolin for useful discussions, and I am particularly
grateful to J. Banavar for a careful critique of this paper.  
I would also like to thank the U.S. Department of Energy for financial
support, and the Alexander von Humboldt foundation for an award.


\end{document}